\documentclass[aps,pre,showpacs,twocolumn]{revtex4}

\usepackage[dvips]{graphicx}
\usepackage{amsmath}
\usepackage{amssymb}
\usepackage{bm}

\begin{document}

\title{The $1/m$ expansion in spin glasses and the de Almeida-Thouless  line}

\author{M.\ A.\ Moore} \email{m.a.moore@manchester.ac.uk}
\affiliation{School of Physics and Astronomy, University of Manchester,
Manchester M13 9PL, UK}

\date{\today}
\begin{abstract}
It is shown by means of a $1/m$ expansion about the large-$m$ limit of the $m$-vector spin glass that the lower critical dimension of the de Almeida-Thouless line in spin glasses is equal to the lower critical dimension of the large-$m$ limit of the $m$-vector spin glass. Numerical studies suggest that this is close to six. 
\end{abstract}

\maketitle
\section{Introduction \label{Introduction}}
Despite the thousands of papers written on the topic of spin glasses over the last four  decades, even their order parameter is still a matter of controversy. The two main rival theories are the replica-symmetry breaking (RSB) theory of Parisi \cite{Parisi}, which is motivated by his exact solution of the Sherrington-Kirkpatrick (SK) model, and the droplet/scaling theory \cite{Fisher-Huse,Bray-Moore, McMillan}. 

Much effort has been devoted to the behavior of spin glasses in a field in an attempt to discriminate between the two theories. According to the RSB picture, there will still be a transition in an applied field $h$, occurring at a temperature $T_c(h)$, which defines the Almeida-Thouless (AT) transition \cite{AT}.  Within the SK model, the AT line can be directly calculated. On the low-temperature side of the line there is a phase with broken replica symmetry, while on the high-temperature or high-field side there is a replica symmetric paramagnetic state. Within the droplet/scaling theory on the other hand there  is no AT line: there is no phase transition in any applied field, just as in a ferromagnet where the addition of a field removes the phase transition.  In this theory the low-temperature phase in zero field is replica symmetric.

There have been  arguments advanced that the RSB picture works for dimensions $d>6$ and that the  droplet/scaling picture only applies when  $d\le6$ \cite{Bray-Roberts, Moore-Bray, Moore}.  On 
this scenario there should be no AT line for $d \le 6$. Numerical simulations to investigate this predicted change of behavior at dimension six are not feasible. Instead investigations have focussed on  whether there is an AT line in three or four dimensional  Ising spin glasses \cite{Jorg, Janus},  with no universally accepted outcome.

In this paper  a new calculational technique is used (at least it is new in the area of spin glasses): the $1/m$ expansion method.  With it a strong argument can be given  that $6$ is indeed the dimension at which the AT line disappears. In particular I argue that the lower critical dimension of the AT line  is likely to be the same as the lower critical dimension for the spin glass phase of the large-$m$ model in zero field. From numerical studies that is known to be around $6$ \cite{Lee-Young, Beyer-Weigel}. 

In the $1/m$ expansion method one studies first the  large $m$, i.e. the $m=\infty$ limit, of the $m$-component vector spin glass ($m=3$ corresponds to the Heisenberg spin glass, $m=1$ is the Ising spin glass) and then proceeds to the construction of the $1/m$-expansion. This procedure was advocated some time ago \cite{Aspelmeier-Moore}. Unfortunately, unlike in the case of ferromagnets, the large-$m$ limit is not analytically tractable, and so carrying out the expansion requires having a good numerical procedure. While numerical work for the large-$m$ limit is much easier than for finite values of $m$ \cite{Morris, Lee-Dhar-Young, BWM}, it is still non-trivial and this is probably the reason why this  paper is effectively the first paper on the $1/m$ expansion for spin glasses, although a start  was made in Ref. \cite{Aspelmeier-Moore}. 

The large-$m$, $m=\infty$, model itself is rather unusual. Its upper critical dimension is eight \cite{GBM}.  That of finite $m$ spin glasses is six \cite{Harris-Lubensky-Chen}. Determination of  its lower critical dimension has to date not been achieved by analytical  studies. In her first study   Viana \cite{Viana} suggested it was $8$, but  a subsequent study revised that to 14 \cite{Viana-Villarreal}.  The conclusion we would draw is that further studies are  needed to obtain the lower critical dimension analytically.  We think it is 6,   but we would acknowledge that this has yet to be properly established. 

What aroused my interest in working on the $1/m$-expansion was a result of Sharma and Young \cite{SY} for the $m$-component SK model in the presence of an $m$-component random field. They  showed that at $T=0$,  the AT line hits the $h$-axis at a finite value, $h_{AT}$, where 
\begin{equation}
h_{AT}^2=\frac{1}{m-2} J^2,
\label{hATSK} 
\end{equation}
provided $m>2$. Thus $h_{AT}^2$ will have a well-defined expansion in $1/m$ whose first terms are 
\begin{equation}
h_{AT}^2=\left[\frac{1}{m}+\frac{2}{m^2} +\cdots \right]J^2.
\label{hATexpansion}
\end{equation}
It occurred to me that it might be possible  to obtain an expression for $h_{AT}^2$ outside the SK limit in terms of correlation functions of the large-$m$ model, by means of a $1/m$ expansion.

 In the large-$m$ limit itself there is no AT line, i.e.  there is no transition in the presence of a $m$-component random field, even in the SK limit.  There is a transition in zero field and the  low-temperature phase is replica symmetric  \cite{Aspelmeier-Moore}.  In order to get an AT line one has to go  to order $1/m$.  We show in this paper  that   $h_{AT}$ at $T=0$ is given by
 \begin{equation}
 h_{AT}^2=\frac{1}{mN}\sum_i \frac{1}{\chi_{ii}^2} +O(1/m^2),
 \label{introhAT}
 \end{equation}
 where $\chi_{ii}$ is the local  susceptibility at site $i$ in the large-$m$ limit. In the SK limit, $\chi_{ii}=1/J$,  so our general expression in Eq. ({\ref{introhAT}), which should be valid in any dimension where there is an AT line, is consistent with that of Sharma and Young for the SK model. 
 
Sharma and Young \cite{SY}  found also  that the degeneracy of the fields which go soft at the AT line in the replica field theory of the $m$-component field spin glass model is $n(n-3)/2$,  for all values of $m$, just as for the  Ising spin glass  at its AT line \cite{Bray-Roberts}, This suggests that the AT line in the $m$-vector model is in the same universality class as that of the Ising model and that it will have the same effective field theory as was used in \cite{Bray-Roberts}. Thus although our work is focussed  on the AT line in the $m$-vector spin glass with $m$ large, our conclusions about its lower critical dimension are applicable also to the lower critical dimension of  the AT line of the Ising spin glass.
 
Unfortunately, to use Eq. (\ref{introhAT}), one needs information on the $\chi_{ii}$ of the large-$m$ model in dimension $d> 6$, which would be very challenging to obtain  numerically. Fortunately, there are arguments which enable one to estimate their magnitude: $\chi_{ii}\sim 1/T_c$, where $T_c$ is the transition temperature of the large-$m$ model in dimension $d$. Then
\begin{equation}
h_{AT}^2 \sim T_c(d)^2/m.
\end{equation}
We suspect that the lower critical dimension of the large-$m$ limit to be six, so $T_c$ should go  to zero as $d \rightarrow 6$. Then the AT field $h_{AT}^2$ should also go to zero in six dimensions:  the lower critical dimension of the large-$m$ limit and the AT line should be the same.

  In Sec. \ref{formalism} the model which we shall study is defined and the formalism  for carrying out the $1/m$ expansion is set up. It is a saddle-point procedure and the leading terms for the large $m$ limit are also given in that section.  The formalism is complicated because of the need to obtain the spin glass susceptibility, which is the susceptibility which diverges at the AT line. In Sec. \ref{oneloop} the one loop calculation is done for the energy of the system and the spin glass susceptibility at  zero temperature, and from the latter we  obtain  the remarkably simple formula for the AT field at zero temperature given in Eq. (\ref{introhAT}).  Our arguments on the form of $\chi_{ii}$ are given in Sec. \ref{finited}. We conclude with a discussion in Sec. \ref{Discussion}.  Finally in the Appendix, it is explained why the problem of understanding the AT line is so hard: it is because it is fundamentally a non-perturbative problem.
  
\section{Formalism for the large $m$ limit \label{formalism}}
We shall study the $m$-component spin version of the Edwards-Anderson spin glass model in an $m$-component random field. This has Hamiltonian
\begin{equation}
\mathcal{H}=-\frac{1}{2}\sum_{i,j}^{N} J_{ij} \sum _{\mu =1}^{m} S_{i\mu}S_{j\mu} -\sum_{i=1}^{N} \sum_{\mu =1}^{m} h_{i\mu} S_{i\mu},
\label{EAHamiltonian}
\end{equation}
where the random field components $h_{i\mu}, \mu =1,2,\cdots, m$, are each drawn from  a Gaussian distribution of zero mean and variance $h^2$ so that $\overline{h_{i\mu}^2}=h^2$. The $m$-component vector spins $\mathbf{S}_i$ are chosen to be of fixed length: 
\begin{equation}
\sum_{\mu=1}^m S_{i\mu}^2=m, \hspace{0.5cm} i=1,2,\cdots, N.
\label{constraint}
\end{equation}
The spins will be taken to sit on the sites of some $d$-dimensional lattice where the sites are labelled $i=1,2,\cdots, N$. The coupling between the spins, $J_{ij}$, could have  any distribution which leads to a spin glass phase. The form of this distribution is not crucial to what follows but it is convenient to imagine that it is between nearest neighbor sites on the lattice  and has a symmetric Gaussian distribution of width $J$ and zero mean. The SK limit is when the interactions are between all pairs of spins: the width of the distribution has then to be set to $J/\sqrt{N}$ in order to make the energy in the thermodynamic limit extensive.

Because we want to study the AT line, we need to calculate the replicon  (spin glass) susceptibility $\chi_R$. It is the divergence of $\chi_R$  as the field $h$ is reduced
while keeping the temperature $T$ fixed which determines the location of  the AT line $h(T)$ \cite{AT}. In order to calculate the replicon susceptibility it is convenient to employ two copies of the system: the Hamiltonian then is
\begin{eqnarray}
\mathcal{H}[\mathbf{S},\mathbf{T}]&=-&\frac{1}{2}\sum_{i,j}^{N} J_{ij}\sum _{\mu =1}^{m} (S_{i\mu}S_{j\mu}+T_{i\mu}T_{j\mu}) \nonumber \\
&-&\sum_{i=1}^{N}\sum_{\mu =1}^{m} h_{i\mu} (S_{i\mu}+T_{i\mu}),
\label{copy}
\end{eqnarray}
where the fixed length spins $\mathbf{T}_i$ are again such that $\sum_{\mu}^mT_{i\mu}^2=m$ at all sites $i$. The Edwards-Anderson order parameter $q$ can be expressed in terms  of $S_{i\mu}$ and $T_{i\mu}$:
\begin{equation}
q \equiv \frac{1}{Nm}\sum_{i}^N\sum_{\mu=1} ^m \overline{\langle S_{i\mu}\rangle^2}=\frac{1}{Nm}\sum_i^N\sum_{\mu=1}^m\overline{\langle S_{i\mu}T_{i\mu}\rangle}.
\label{qEAdef}
\end{equation}
(The overline indicates the average over the random field distribution. We shall have no need to average over the bond distribution of the couplings $J_{ij}$).
Similarly  the replicon susceptibility $\chi_R$ is 
\begin{eqnarray}
&\chi_R&\equiv \frac{1}{Nm} \beta^2 \sum_{i,\mu,j,\nu}\overline{[\langle S_{i\mu}S_{j\nu}\rangle-\langle S_{i\mu}\rangle \langle S_{j\nu}\rangle]^2} =\frac{1}{Nm} \beta^2   \nonumber \\
&\times&\sum_{i,\mu,j,\nu}\overline{[\langle S_{i\mu}S_{j\nu}\rangle-\langle S_{i\mu}\rangle \langle S_{j\nu}\rangle][\langle T_{i\mu}T_{j\nu}\rangle-\langle T_{i\mu}\rangle \langle T_{j\nu}\rangle]},
\nonumber
\end{eqnarray}
where here $\beta=1/T$.

To handle the constraints imposed by having fixed length spins, it is convenient to make use of the representation \cite{Bray-Moore82, Aspelmeier-Moore}
\begin{equation}
\delta(m-\sum_{\mu =1}^m S_{i\mu}^2)=\int_{-i\infty}^{\i \infty} \frac{\beta d H_i}{4 \pi} \exp[\beta(m-\sum_{\mu=1} ^m S_{i\mu}^2)H_i/2].
\nonumber
\end{equation}
The partition function of  two copies, $Z^2$, where $Z$ is the partition function of a single copy derived from the Hamiltonian of Eq. (\ref{EAHamiltonian}),  can  be written
\begin{eqnarray}
&&Z^2=\int \prod \frac{\beta dH_i}{4 \pi} \int \prod\frac{\beta dK_i}{4 \pi} \int_{-\infty}^{\infty}\prod dS_{i\mu}\int_{-\infty}^{\infty}\prod dT_{i\mu} \nonumber \\
&&\exp \Bigg \{-\beta\mathcal{H}[\mathbf{S},\mathbf{T}]+
\nonumber \\
&& \sum_i\left[\beta(m-\sum_{\mu=1} ^m S_{i\mu}^2)H_i/2+\beta(m-\sum_{\mu=1} ^m T_{i\mu}^2)K_i/2 \right] 
\Bigg \}.  
\nonumber
\end{eqnarray}

In order  to calculate the free energy and correlation functions such as $\chi_R$, averaged over the random fields, it is necessary to use the replica trick and obtain $\overline{Z^{2n}}$.  As usual $n$ has to be set zero at the end of the calculation. To this end one attaches replica labels so that  $H_i$, $K_i$,  $S_{i\mu}$ and $T_{i\mu}$   become $H_i^{\alpha}$, $K_i^{\alpha}$, $S_{i\mu}^{\alpha}$ and $T_{i\mu}^{\alpha}$, where $\alpha=1,2, \cdots, n$. Then 
\begin{eqnarray}
&&Z^{2n}=\int \prod \frac{\beta dH_i^{\alpha}}{4 \pi} \int \prod\frac{\beta dK_i^{\alpha}}{4 \pi} \int \prod dS_{i\mu}^{\alpha}\int \prod dT_{i\mu}^{\alpha}  \nonumber \\
& &  \exp\Bigg\{\frac{\beta}{2}\sum_{\alpha=1}^n \sum_{i,j}^{N} J_{ij}\sum _{\mu =1}^{m} (S_{i\mu}^{\alpha}S_{j\mu}^{\alpha}
+T_{i\mu}^{\alpha}T_{j\mu}^{\alpha}) \nonumber \\
&&+\beta \sum_{i=1}^{N}\sum_{\mu =1}^m h_{i\mu} \left[\left(\sum_{\alpha=1}^n S_{i\mu}^{\alpha}\right)+\left(\sum_{\alpha=1}^n T_{i\mu}^{\alpha}\right)\right] +\nonumber \\
&&\sum_{i,\alpha}\left[\beta(m-\sum_{\mu=1} ^m S_{i\mu}^{\alpha 2})H_i^{\alpha}/2+
\beta(m-\sum_{\mu=1} ^m T_{i\mu}^{\alpha 2})K_i^{\alpha}/2\right] \Bigg\}.
 \nonumber
\end{eqnarray}
On averaging over the random fields $h_{i\mu}$
\begin{eqnarray}
&&\overline{Z^{2n}}=\int \prod \frac{\beta dH_i^{\alpha}}{4 \pi} \int \prod\frac{\beta dK_i^{\alpha}}{4 \pi}  \int \prod dS_{i\mu}^{\alpha}\int \prod dT_{i\mu}^{\alpha}  \nonumber \\ 
&&\exp \Bigg \{ \frac{\beta}{2}\sum_{\alpha=1}^n\sum_{i,j}^{N} J_{ij}\sum _{\mu =1}^{m} (S_{i\mu}^{\alpha}S_{j\mu}^{\alpha}
+T_{i\mu}^{\alpha}T_{j\mu}^{\alpha})\nonumber \\
&& +\sum_{i=1}^{N}\sum_{\mu =1}^m \Bigg[ \frac{\beta^2 h_D^2}{2} \left (\left (\sum_{\alpha=1}^n S_{i\mu}^{\alpha}\right)^2
+\left(\sum_{\alpha=1}^n T_{i\mu}^{\alpha}\right)^2 \right) \nonumber \\
&&+\beta^2 h_M^2 \left(\sum_{\alpha=1}^n S_{i\mu}^{\alpha}\right)\left(\sum_{\beta=1}^n T_{i\mu}^{\beta}\right )\Bigg] +\nonumber \\
&&\sum_{i,\alpha}\left[\beta(m-\sum_{\mu=1} ^m S_{i\mu}^{\alpha 2})H_i^{\alpha}/2+
\beta(m-\sum_{\mu=1} ^m T_{i\mu}^{\alpha 2})K_i^{\alpha}/2\right]  \Bigg \}.
\nonumber
\end{eqnarray}
We have introduced the variables $h_D$ and $h_M$ in order to generate various correlation functions from differentiating the free energy, $F/T=-\ln \overline{Z^{2n}}$, but after differentiating with respect to them one sets at the end of the calculation
\begin{equation}
h_D=h_M =h.
\label{hMhD}
\end{equation}
The free energy $F$ then of the replicated system  is just $2n$ times the free energy of the unreplicated system. For example, given $F$ one can obtain the Edwards-Anderson order parameter $q$ as defined in 
Eq. (\ref{qEAdef}) via
\begin{eqnarray}
-\frac{\partial (F/T)}{\partial (\beta^2 h_M^2)}&=&\sum_{i,\mu} \langle \left(\sum_{\alpha=1}^n S_{i\mu}^{\alpha}\right)\left(\sum_{\beta=1}^n T_{i\mu}^{\alpha}\right )\rangle  \nonumber\\
&=&n^2 \sum_{i,\mu} \langle S_{i\mu}\rangle^2= n^2 N m q.
\label{repqdef}
\end{eqnarray}
Second derivatives with respect to $\beta^2 h_M^2$ yield the replicon susceptibility:
\begin{eqnarray}
&&-\frac{\partial^2(F/T)}{\partial(\beta^2 h_M^2)^2}
=\sum_{i,\mu,j,\nu} \bigg[ \langle \sum_{\alpha=1}^n S_{i\mu}^{\alpha} \sum_{\beta=1}^n T_{i\mu}^{\beta}\sum_{\gamma=1}^n S_{j\nu}^{\gamma} \sum_{\delta=1}^n T_{j\nu}^{\delta} \rangle   \nonumber \\
&&\quad  \quad \quad \quad \quad -\langle \sum_{\alpha=1}^n S_{i\mu}^{\alpha} \sum_{\beta=1}^n T_{i\mu}^{\beta} \rangle \langle \sum_{\gamma=1}^n S_{j\nu}^{\gamma} \sum_{\delta=1}^n T_{j\nu}^{\delta} \rangle \bigg] \nonumber \\
&&=n^2\sum_{i,\mu,j,\nu}[\langle S_{i\mu} S_{j\nu}\rangle-\langle S_{i\mu}\rangle \langle S_{j\nu}\rangle][\langle T_{i\mu} T_{j\nu}\rangle-\langle T_{i\mu}\rangle \langle T_{j\nu}\rangle]\nonumber\\
&& +{\rm O} (n^3)=n^2 Nm \chi_R/\beta^2 \quad {\rm as} \quad  n \rightarrow 0.
\label{chiviaF}
\end{eqnarray}
We thus will need to calculate the free energy $F$ to order $n^2$ to get $\chi_R$.

The next step is to integrate out the spin variables $S_{i\mu}^{\alpha}$ and $T_{i\mu}^{\alpha}$. This is readily done using a cumulant expansion. It is convenient to let the replica indices on $H_i^{\alpha}$  run $1,2,\cdots, 2n$, and set $K_i^{\alpha}=H_i^{\alpha}$ when $\alpha=n+1,n+2, \cdots, 2n$. We shall define
\begin{equation}
A_{ij}=H_i^{\alpha}-J_{ij},
\label{Adef}
\end{equation} 
and call its inverse
\begin{equation}
\chi_{ij}^{\alpha}=[A^{-1}]_{ij}.
\label{chidef}
\end{equation}
Then the remaining integrals for $\overline{Z^{2n}}$ are
\begin{equation}
\overline{Z^{2n}}=\int_{-i \infty}^{\i \infty} \prod_{\alpha=1}^{2n} \frac{\beta dH_i^{\alpha}}{4 \pi} \exp[m S(\{H_i^{\alpha}\})],
\label{saddleint}
\end{equation}
where, up to the third cumulant (and terms of order $n^3$)
\begin{eqnarray}
&&S(\{H_i^{\alpha}\})=\frac{\beta}{2}\sum_{i,\alpha=1}^{2n}[H_i^{\alpha}+\frac{1}{\beta} \ln \det (\chi^{\alpha}_{ij}/\beta) +h_D^2 \chi_{ii}^{\alpha}]  + \nonumber \\
&&\frac{\beta^2}{4}\left(h_D^4\sum_{\alpha,\beta=1}^{n}+h_D^4\sum_{\alpha,\beta=n+1}^{2n}+2 h_M^4
\sum_{\alpha=1}^n\sum_{\beta=n+1}^{2n}\right) \chi_{ij}^{\alpha}  \chi_{ji}^{\beta} \nonumber \\
&&+\frac{\beta^3}{6}\Bigg(h_D^6 \sum_{\alpha,\beta,\gamma=1}^n+h_D^6 \sum_{\alpha,\beta,\gamma=n+1}^{2n}+\nonumber \\
&&3 h_D^2 h_M^4 \sum_{\alpha,\beta=1}^n \sum _{\gamma=n+1}^{2n}+3 h_D^2 h_M^4\sum_{\alpha=1}^n\sum_{\beta,\gamma=n+1}^{2n}\Bigg)  \chi_{ij}^{\alpha}\chi_{jk}^{\beta}\chi_{ki}^{\gamma},\nonumber
\end{eqnarray}
where we have adopted the notation that repeated site indices are to be summed from $1$ to $N$.

Because of the factor of $m$, which we are taking to be large in Eq. (\ref{saddleint}), the integrals over $H_i^{\alpha}$ can be done by the method of steepest descents. The leading term can be found from the solutions of 
\begin{equation}
\frac{\partial S}{\partial H_{i}^{\alpha}}=0, \,\,\, {\rm for} \,\,\, i=1,\cdots, N,\,\,\, {\rm and} \,\,\, \alpha =1,\cdots, n.
\nonumber
\end{equation}
There is a solution of these $Nn$ set of equations which has replica symmetry, $H_i^{\alpha} =H_i$ for $\alpha=1,\cdots,n$. The values of $H_i$ can be obtained by solving the $N$ saddle-point equations,
\begin{eqnarray}
 \beta-\chi_{ii}-\beta h_D^2 \sum_k \chi_{ik} \chi_{ki}- n\beta^2(h_D^4+h_M^4)\sum_{k,l} \chi_{ik}\chi_{kl}\chi_{li}\nonumber \\
-n^2 \beta^3(h_D^6+3 h_D^2 h_M^4)\sum_{k,l,m}\chi_{ik}\chi_{kl}\chi_{lm}\chi_{mi}=0. \hspace{1.0cm}
\label{saddle}
\end{eqnarray}
We shall work in the low-temperature limit $T \rightarrow 0$ or $\beta \rightarrow \infty$, which permits a number of simplifications. In this limit, the term in  $\chi_{ii}$, which is finite at $T=0$, can be dropped  as it is negligible in comparison to the terms with  factors of $\beta$. In the limit of $n\rightarrow 0$,  $H_i\rightarrow H_i^{0}$ which can be determined by solving the $N$ equations
\begin{equation}
1 =h_D^2 \sum_k\chi_{ik}\chi_{ki}.
\nonumber
\end{equation}
Using the saddle-point solution the free energy to order $n^2$ in the large m limit as $T\rightarrow 0$  is then 
\begin{equation}
F/T=-\beta nm \sum_i H_i-\frac{\beta^2 n^2m  N}{2h_D^2} (h_D^4+h_M^4).
\label{largemF}
\end{equation}
It follow on using Eq. (\ref{repqdef}) that $q=h_M^2/h_D^2$ which equals $1$ on setting $h_D=h=h_M$. This is what  would be expected at zero temperature. The replicon susceptibility can be calculated using Eq. (\ref{chiviaF}) and is given by $\chi_R=1/h_D^2=1/h^2$.

It will be important for what follows at one-loop order that we know the leading changes which a finite value of $n$ would make to the $H_i$. If we set $H_k\rightarrow H_k^{(0)}+\Delta H_k$,
then 
\begin{equation}
\chi_{ij}\rightarrow \chi_{ij}^{(0)}-\sum_{k}\chi_{ik}^{(0)}\Delta H_k \chi_{kj}^{(0)}.
\label{deltachi}
\end{equation}
From  Eq. (\ref{saddle}), to order $n$,
\begin{equation}
\sum_k \chi_{ik}^2=\frac{1}{h_D^2}-\beta n(h_D^4+h_M^4) \frac{\Delta_i}{h_D^2},
\nonumber
\end{equation}
where 
\begin{equation}
\Delta _i=\sum_{k,l} \chi_{ik}^{(0)} \chi_{kl}^{(0)} \chi_{li}^{(0)}.
\label{triangle}
\end{equation}
By substituting for $\chi_{ik}$ using   Eq. (\ref{deltachi})  one has  to order $n$
\begin{equation}
-2 \sum_{k}\chi_{ik}^{(0)}\Delta H_k \chi_{kj}^{(0)} \chi_{ji}^{(0)}=-\beta n(h_D^4+h_M^4) \frac{\Delta_i}{h_D^2}.
\nonumber
\end{equation}
which has solution 
\begin{equation}
\Delta H_k=\Delta_H=\frac{\beta n}{2h_D^2} (h_D^4+h_M^4), \quad k=1,\cdots, N.
\label{shiftn}
\end{equation}
$S(\{H_i\})$ goes like $n$ as $n\rightarrow 0$. Hence changes of order $n$ in  $H_i$ will only contribute terms of order $n^3$ to S.  This is because $S$ is stationary with respect to variations of $H_i$. 

We shall need a special case of Eq. (\ref{deltachi}) in the one-loop order calculation:
\begin{eqnarray}
\chi_{ii}^{(1)}&=&\chi_{ii}^{(0)}-\sum_k\chi_{ik}^{(0)}\Delta_H \chi_{ki}^{(0)}\nonumber \\
&=&\chi_{ii}^{(0)}-\frac{\beta n}{2h_D^4} (h_D^4+h_M^4).
\label{chi1}
\end{eqnarray}

\section{One-Loop order contribution \label{oneloop}}
The one-loop order contribution  to $F$ comes from the Gaussian fluctuations about the saddle-point solution $H_i^{\alpha}=H_i$. Writing
$\delta H_i^{\alpha}=H_i^{\alpha}-H_i$, the terms in the expansion of $S$ in Eq. (\ref{saddleint})  to quadratic order in $\delta H_i^{\alpha}$ are 
\begin{equation}
\frac{1}{2}\sum_{ij} \sum_{\alpha,\beta=1}^{2n} \delta H_i^{\alpha} M_{\alpha \beta}(ij) \delta H_j^{\beta},
\nonumber
\end{equation}
where the matrix
\begin{equation}
M_{\alpha \beta}(ij)=\frac{\partial^2 S}{\partial H_i^{\alpha} \partial H_j^{\beta}},
\nonumber
\end{equation}
is evaluated at the saddle-point $H_i^{\alpha}=H_i$.
\begin{eqnarray}
&&M_{\alpha \beta}(ij)=
\delta_{\alpha\beta} C_{ij}\nonumber \\
&& +\beta^2 h_D^4 D_{ij}+ 2 n \beta^3(h_D^6+h_D^2 h_M^4) P_{ij}, \quad 1\le \beta \le n,\nonumber \\
&& +\beta^2 h_M^4 D_{ij}+4 n \beta^3 h_D^2 h_M^4 P_{ij},\quad n+1 \le \beta \le 2n, \nonumber
\end{eqnarray}
provided  $\alpha \le n$. The case of $\alpha \ge n+1$ is   obtained by switching around the two forms dependent on  $\beta$. Here 
\begin{eqnarray}
D_{ij}&=&\sum_{k,l} \chi_{ik}\chi_{il} \chi_{kj}\chi_{lj}, \quad
P_{ij}=\sum_{k,l,m} \chi_{ik} \chi_{kl}\chi_{lj}\chi_{jm}\chi_{mi}, \nonumber \\
S_{ij}&=&\sum_{k,l} \chi_{ij}\chi_{jk} \chi_{kl} \chi_{li}, \quad
Q_{ij}=\sum_{k,l,m} \chi_{ij} \chi_{jk}\chi_{kl}\chi_{lm}\chi_{mi}, \nonumber \\
C_{ij}&=&\chi_{ij}^2+2 \beta h_D^2 \chi_{ij} \sum_k \chi_{ik}\chi_{kj}+  2 n \beta^2 (h_D^4+ h_M^4) S_{ij} \nonumber \\ 
&+&2 n^2 \beta^3(h_D^6+3 h_D^2 h_M^4) Q_{ij}. \nonumber 
\end{eqnarray}
The matrix $M_{\alpha\beta}(ij)$ is a $2n \times 2n$ matrix in replica space with $C_{ij}$ along the diagonal. In the two $n\times n$ blocks near the diagonal the elements are all $\beta^2h_D^4D_{ij}+2 \beta^3(h_D^6+h_D^2 h_M^4)P_{ij}$, while in the  two $n\times n$ blocks off the diagonal the elements are all $\beta^2 h_M^4D_{ij}+4 \beta^3 h_D^2 h_M^4 P_{ij}$. Such a matrix has a  $2n-2$  fold degenerate eigenvalue equal to $C_{ij}$ and two non-degenerate eigenvalues
\begin{eqnarray}
U_{ij}=C_{ij}+n\beta^2(h_D^4-h_M^4)D_{ij}+2n^2 \beta^3(h_D^6-h_D^2 h_M^4)P_{ij}, \nonumber \\
V_{ij}=C_{ij}+n\beta^2(h_D^4+h_M^4)D_{ij}+2 n^2 \beta^3(h_D^6+3 h_D^2 h_M^4)P_{ij}. \nonumber
\end{eqnarray}
If we denote the one-loop contribution to $\overline{Z^{2n}}$ by $\exp S_1$, 
\begin{equation}
S1=-\frac{1}{2}\left[(2n-2) \ln \det C_{ij}+\ln \det U_{ij}+\ln\det V_{ij}\right]. \nonumber
\end{equation}
To make analytical progress we shall work in the limit $T \rightarrow 0$. Furthermore we will find that if we keep $h_D^2$ and $h_M^2$ small, of $O(J^2/m)$, the expected magnitude of the field at the AT line, then substantial simplifications are possible. 

Let us begin by simplifying $C_{ii}$ with the aid of  Eq. (\ref{saddle}). It becomes 
\begin{equation}
C_{ii}=2(\beta-\chi_{ii})\chi_{ii}, \nonumber 
\end{equation}
 if one uses the identities $S_{ii}=\chi_{ii} \sum_{k,l} \chi_{ik}\chi_{kl}\chi_{ki}$and 
$Q_{ii}=\chi_{ii}\sum_{k,l,m}\chi_{ik}\chi_{kl}\chi_{lm}\chi_{mi}$.
As $T\rightarrow 0$, $\chi_{ii}$ remains finite, so 
$C_{ii}\rightarrow 2 \beta \chi_{ii}$. 

The off-diagonal terms in $C_{ij}$ with $i \ne j$ and $n=0$ are in the $T\rightarrow 0$ limit  $2 \beta h_D^2 \chi_{ij}\sum_k \chi_{ik}\chi_{kj}$. Thus in the limit we shall work, when $h_D^2 \sim J^2/m$ they are negligible compared to the diagonal terms and we can write $C_{ij}=2 \beta \chi_{ii}\delta_{ij}$. With this approximation, it is straightforward to evaluate $S1$ to $O(n^2)$.
\begin{eqnarray}
&&S1=-n \sum_i \ln (2\beta \chi_{ii}) -\frac{1}{2} n\beta h_D^4 \sum_i D_{ii}/\chi_{ii} \nonumber \\
&& - n^2 \beta^2(h_D^6+h_D^2 h_M^4) \sum_i P_{ii}/\chi_{ii} \nonumber \\
&&+\frac{1}{8} n^2 \beta^2  
(h_D^8+h_M^8)\sum_{i,j}\frac{1}{\chi_{ii}} D_{ij}^2\frac{1}{\chi_{jj}}.               
\nonumber
\end{eqnarray}
For the terms linear in $n$ we have to use for  $D_{ii}$ its form correct  to order $n$ to get the $n^2$ terms in $S1$:
\begin{equation}
D_{ii}=\left(\sum_k\chi_{ik}^2\right)^2=\frac{1}{h_D^4}[1-2 n \beta (h_D^4+h_M^4)\Delta_i],
\nonumber 
\end{equation}
and also use Eq. (\ref{chi1}) for $\chi_{ii}$.
Now
\begin{equation}
P_{ii}=\sum_m \chi_{im}^2\sum_{k,l}\chi_{ik}\chi_{kl}\chi_{li}=\frac{1}{h_D^2} \Delta_i,
\nonumber
\end{equation}
and  the double sum
\begin{equation}
\sum_{i,j}\frac{1}{\chi_{ii}} D_{ij}^2\frac{1}{\chi_{jj}}\approx \frac{1}{h_D^8}\sum_i \frac{1}{\chi_{ii}^2},
\nonumber
\end{equation}
as for small $h_D$ it is dominated by the term with $i=j$. The term in $S1$ equal to $-n \sum_i \ln (2\beta \chi_{ii})$ is subdominant as $T \rightarrow 0$.  Assembling all the terms together the leading terms up  to one-loop order are
\begin{eqnarray}
&&F/T = -\beta n m \sum_i H_i + \frac{1}{2}\beta n  \sum_i \frac{1}{\chi_{ii}} \nonumber \\
&& -\frac{\beta^2 n^2m  N}{2h_D^2} (h_D^4+h_M^4)\nonumber \\
&&+\frac{\beta^2 n^2 (h_D^8+2 h_D^4 h_M^4 - h_M^8)}{8 h_D^8}\sum_i\frac{1}{\chi_{ii}^2}.
\label{final}
\end{eqnarray}
In this equation, $\chi_{ii}$  and $H_i$ are  at their  $n \rightarrow 0$ values. 

By using Eq. (\ref{repqdef}) one can confirm that this expression for $F/T$ still implies that $q=1$. By using Eq. (\ref{chiviaF}) the replicon susceptibility is, after setting $h_D=h=h_M$,
\begin{equation}
\chi_R= \frac{1}{h^2}+\frac{1}{m h^4}\frac{1}{N} \sum_i\frac{1}{\chi_{ii}^2}.
\label{chiRfinal}
\end{equation}
Terms which have been neglected, like the terms of $C_{ij}$ with $i \ne j$ give rise to subdominant corrections of order $1/(m h^2)$ in $\chi_R$ which are negligible  compared to the second term in Eq. (\ref{chiRfinal}) provided we are only  interested in values of $h^2 \sim J^2/m$.

In the SK limit of the large m model, $\chi_{ii}$ has no site dependence in the limit $N\rightarrow \infty$ and at $T=0$ equals $1/(J^2+h^2)^{1/2}$. If we take $h^2$ to be small, say, of order $J^2/m$, then for the SK model Eq. (\ref{chiRfinal}) reduces to
\begin{equation}
\chi_R=\frac{1}{h^2}+\frac{J^2}{mh^4} +\cdots  \approx \frac{1}{h^2-h_{AT}^2},
\label{simplepole}
\end{equation}
with $h_{AT}^2 =J^2/m$. This result is therefore in perfect agreement with that of Sharma and Young \cite{SY} for the value of the critical field at $T=0$. For all $d>6$ one is in the classical regime where the exponent $\gamma$  characterizing the divergence of $\chi_R$ as one approaches the AT line is equal to $1$, a feature which is captured by the simple pole expression of Eq. (\ref{simplepole}).

Outside the SK limit, the expression for the AT field at $T=0$ generalizes to
\begin{equation}
h_{AT}^2=\frac{1}{mN}\sum_i\frac{1}{\chi_{ii}^2}.
\label{chifinal}
\end{equation}
We shall discuss the consequences of this for the properties of spin glasses in Sec. \ref{finited}.

The ground state energy $E$ of the system is obtained from Eq. (\ref{final})  by dividing the linear term in $n$ of $F$ by the total number of replicas $2n$, so 
\begin{equation}
E/m=-\frac{1}{2} \sum_i H_i+\frac{1}{4m}\sum_i\frac{1}{\chi_{ii}}.
\label{energy}
\end{equation}
In the SK limit, $H_i$ is site independent and
\begin{equation}
H_i= \frac{J^2}{\sqrt{J^2+h^2}}+\sqrt{J^2+h^2}.
\nonumber
\end{equation}
(Solution of the $m=\infty$ SK spin glass model is aided by the observation of de Almeida \textit{et al.} \cite{AJKT} that it is replica symmetric and has the same free energy as the spherical spin glass, \cite{KTJ}, although the physics of the two are quite different \cite{Aspelmeier-Moore}).

The $1/m$ expansion done in this paper will only have utility in the region of fields and temperatures 
above the AT line. Nevertheless, provided that we take $h^2>h_{AT}^2$, but still of order $J^2/m$, it is possible to obtain  the ground state energy at $h=0$ correct to order $1/m$, as can be seen by expanding out the expression for $H_i$ in a power series in $h^2$: $H_i=2J+h^4/4 \cdots$. Hence in the SK limit, the ground state energy is given by
\begin{equation}
E/m=-J+\frac{J}{4m} +O(J/m^2).
\end{equation}
Setting $m =1$, the ground state energy per spin of the Ising spin glass  is estimated to be  $-0.75 J$. It is actually equal to $-0.763 \dots J$, \cite{Parisi}, suggesting that the $1/m$ expansion might have utility even for Ising systems!

In general one would expect that the $1/m$ expansion would only work when the state about which one is expanding, the large-$m$ limit, is similar to the state at finite values of $m$. The low-temperature phase of the SK model in a random field at finite  values of $m$ has full replica symmetry breaking and  is quite different to the paramagnetic, replica symmetric state of the large-$m$ limit. So it is rather surprising that putting $m=1$ gives quite a good result for the ground state energy of the SK model. However, in say two dimensions, there is certainly no AT line at finite values of $m$, and one then could hope that the $1/m$ expansion for the energy should work well. It should be easy to obtain by numerical methods  the $H_i$ and the $\chi_{ii}$  in both two and three dimensions which have to be  inserted into Eq. (\ref{energy}) in order to check this out.

\section{The AT field in finite dimensions \label{finited}}
In order to calculate $h_{AT}^2$  using Eq. (\ref{chifinal}) one needs information on the $\chi_{ii}$, which are, alas, unknown.  We can only make progress by examining their likely dependence on the value of $T_c$. 

 So long as there is a finite temperature transition  in the large-$m$ limit model, $\chi_{ii}$ will be finite as $h$ goes to zero and it is its value at $h=0$  and at $T=0$ which should be used in Eq. (\ref{chifinal}).  At $h=0$, the saddle point equations simplify to $\beta=\chi_{ii}$ \cite{Aspelmeier-Moore} when $T>T_c$.  For $T<T_c$,
\begin{equation}
\chi_{ii}=\beta (1-q_i),
\end{equation}
where 
\begin{equation}
q_i=\frac{1}{m}\sum_{\mu=1}^m \langle S_{i\mu}\rangle^2.
\nonumber
\end{equation}
 Below $T_c$, there is a Bose-Einstein like condensation into a large number of modes, $m_0$. For  the SK limit $m_0 \sim N^{2/5}$  \cite{Aspelmeier-Moore}.  Each component $\mu$  of $\langle S_{i\mu} \rangle$ is proportional to a different one of the $m_0$ nearly null eigenvectors of the $A_{ij}$ matrix.  $q_i$ is the sum of $m_0$ terms 
 \begin{equation}
 q_i=\frac{1}{m_0}\sum_{\mu=1}^{m_0} \langle S_{i\mu}\rangle^2.
 \nonumber
 \end{equation}
 The site average of $q_i$ is the Edwards-Anderson order parameter q.
 
In the SK limit $q=1-T/T_c$, so $\chi=\beta(1-q)=1/T_c$ for all $T<T_c$. Outside the SK limit, $q$ is not known precisely, but as $T\rightarrow 0$ it would seem natural to expect that it goes like $q=1-aT/T_c$, where $a$ is a constant. This is certainly consistent with the results in Ref. \cite{BWM} (albeit on the long-range one-dimensional long-range version of the large $m$ model).  Similarly one would expect that the low-temperature form of the $q_i$ should be $q_i=1-a_iT/T_c $. Then $\chi_{ii}=a_i/T_c$ at $T=0$ and
\begin{equation}
h_{AT}^2=\frac{k(d)}{m } T_c^2 ,\quad k(d)=\frac{1}{N} \sum_i \frac{1}{a_i^2}.
\label{k(d)}
\end{equation}
Not much is known about the coefficients $a_i$.  In the SK limit, $\chi_{ii}=1/J =1/T_c$ so $a_i=1$ at all sites. Thus in that limit $k=1$. We shall assume that $k(d)$ remains finite as $d \rightarrow 6$.

Normally, as a system approaches its lower critical dimension its transition temperature falls to zero. As we suspect that the lower critical dimension of the large-$m$ limit is $6$, it would follow then from Eq. (\ref{k(d)})  that the AT line should also disappear as $d\rightarrow 6$.

The dependence of $T_c$ on  dimensionality  $d$ can be elucidated further if the McMillan RG equation \cite{McMillan} applies  to the large-$m$ model. Certainly it seems to work in the one-dimensional version of the model studied in Ref. \cite{BWM}. McMillan argued that the flow of the temperature $T$ under a renormalization group transformation near the lower critical dimension where $\theta$ is small was as 
\begin{equation}
\frac{dT}{d\ln L}=-\theta T+cT^3/J^2+ \cdots,
\end{equation}
and c is a constant of order one.  Provided that $\theta$ is positive, this has a fixed point when $T_c^2 =\theta J^2 /c$. Near the lower critical dimension $\theta$ changes sign, so let us write it as $\theta \sim (d-6)$. Then close to six dimensions,
\begin{equation}
h_{AT}^2  \sim \frac{d-6}{m } k(d) J^2.
\label{hATfinal}
\end{equation}

Thus according to this  calculation, at one-loop order the AT line should not exist below $6$ dimensions, or more conservatively, the lower critical dimension of the large-$m$ model.

\section{Discussion} \label{Discussion}

The  remaining major task is to show conclusively that $T_c$, the transition temperature in the large-$m$ limit at zero field, does  indeed go to zero as $d \rightarrow 6$ or in other words, that six is the lower critical dimension in the large-$m$ limit. There are intriguing features of the large-$m$ limit which suggest that this problem is not as hard as that of directly determining the lower critical dimension for the AT transition. There is a suggestion in \cite{GBM} that the exponents in the large-$m$ limit can be related to the critical exponents of a problem without disorder in two fewer dimensions, which is an example of dimensional reduction. Dimensional reductions can often  be related to some  supersymmetry, so maybe an elegant demonstration  that six is the lower critical dimension will be possible.

 Assumptions have had to be been made in this calculation because the $\chi_{ii}$ are not known outside the SK limit.    We could make some progress by using  the usual assumptions as to the form of $q$ at low temperatures, but we could not work out the $\chi_{ii}$  explicitly.
Unfortunately, numerical studies of the large-$m$ model near six dimensions  to check the assumed form of $\chi_{ii}$ are unlikely to be practical.

However, it is possible to do numerical work on the one-dimensional version of the model with long-range interactions, as in Ref.  \cite{BWM}. 
Here the interactions between the spins are still random but fall off with a power  of their separation $r_{ij}$: $J_{ij} \sim \epsilon_{ij}/r_{ij}^{\sigma}$. The $\epsilon_{ij}$ are Gaussian random variables of variance $J^2$, while the exponent $\sigma$ can be tuned to mimic various dimensions of the short-range model in $d$ dimensions \cite{BWM}.  In zero field, the SK model corresponds to $\sigma <1/2$, the upper critical dimension 8 of the large-$m$ limit maps to $\sigma=5/8$. The lower critical dimension (which I think is 6) corresponds to $\sigma =3/4$. The numerical work in Ref. \cite{BWM} provides good evidence that $T_c^2 \sim (0.75 - \sigma)J^2$. 

Now I believe that the AT line at any \textit{finite} value of $m$ disappears for $d \le  6$  or values of $\sigma \ge 2/3$ \cite{Moore-Bray}. But our $1/m$ expansion to first order in $1/m$ would predict that $h_{AT}^2$ is finite until $\sigma$ reaches $3/4$.  I suspect that  non-perturbative effects will  remove the AT line in the interval $2/3\le \sigma <3/4$. This will be discussed  in a separate paper on the AT line for large but finite $m$ in the  one-dimensional model \cite{1DAT}.

It is easy to see that non-perturbative effects must be important. Suppose there were an AT line for $2/3 \le \sigma <3/4$.  There is no stable perturbative fixed point  in  the vicinity of the \lq\lq upper critical dimension", i.e. when $\sigma \ge 2/3$, \cite{ Moore-Bray}, indicating that the calculations in this interval have to be non-perturbative. Non-perturbative effects must also be important for $\sigma < 2/3$ in determining the value of $h_{AT}^2$, even though the critical behavior at the AT line is there controlled by the Gaussian fixed point \cite{Moore-Bray}.  Non-perturbative effects must  reduce  $h_{AT}^2$ to zero, probably as some power of  $(2/3 -\sigma)$, so that $h_{AT}^2$ joins smoothly to being zero for $\sigma \ge 2/3$.

For the spin glass model in $d$ dimensions it is a happy coincidence that the lower critical dimension of the AT line and that of the large-$m$ limit both seem to be six. It is likely  though that the same type of non-perturbative effects which must dominate the one-dimensional model for $\sigma$ close to $2/3$ might affect some of our estimates of $h_{AT}^2$ when $d$ is close to  $6$. Our one-loop order calculation has suggested that $h_{AT}^2 \sim T_c^2 A(m,d)$  where  $A(m,d) \sim 1/m$ as $m \rightarrow \infty$.  Contributions from higher orders in the loop expansion would still be expected to go as $T_c^2$, and would just add further  terms in the $1/m$ expansion of $A(m,d)$, as in Eq. (\ref{hATexpansion}) for the SK limit (recall that in this limit $J^2=T_c^2$).  Some of the dimensionality dependence of $h_{AT}^2$ will arise from that of $T_c^2$, which I think  varies as $(d-6)J^2$.  Non-perturbative effects might at fixed $m$ make $A(m,d)$ vanish as some power of $(d-6)$ just as they might produce powers of $(2/3-\sigma)$ in the one-dimensional long-range model. But even without being able to solve the non-perturbative problem, we can still be confident that there is no AT line in less than six dimensions, as $T_c$ will vanish in six  dimensions.

Finally, the $1/m$ expansion procedure is quite general and can be used for any quantity in principle. It would be good to apply it to problems for which there is no analytical work, such as the value of the droplet exponent $\theta$ \cite{Fisher-Huse, Bray-Moore, McMillan}.

\begin{acknowledgments}
I should like to thank Frank  Beyer and Martin Weigel  for re-kindling my interest in the large-$m$ limit. I would also like to thank Peter Young for his useful comments.
\end{acknowledgments}

\appendix*
\section{Why calculating the AT line near six dimensions is hard}
In this Appendix we explain why investigating the behavior of the AT line as $d \rightarrow 6$ is hard for Ising spin glasses.   It is  because it is a non-perturbative problem.

The  Ginzburg-Landau-Wilson free-energy functional  for the Ising
spin  glass written in  terms of  the replica  order parameter
field is \cite{Harris-Lubensky-Chen}
\begin{eqnarray}
&F[\{Q_{\alpha\beta}\}]  =   \int   d^dx\,   \left[\frac{1}{2}r
\sum_{\alpha<\beta}Q_{\alpha\beta}^2
+\frac{1}{2}\sum_{\alpha<\beta}(\nabla
Q_{\alpha\beta})^2\right. \nonumber  \\
&\hspace*{-0.2cm}  +   \left.\frac{w}{6}
\sum_{\alpha<\beta<\gamma}Q_{\alpha\beta}
Q_{\beta\gamma}Q_{\gamma\alpha}  -  h^2
\sum_{\alpha<\beta} Q_{\alpha\beta} + O(Q^4)\right]
\nonumber
\end{eqnarray}
where  $h$  is the  applied  field.  A simple scaling
analysis of the  terms in the functional  shows that
the  natural size  of $h_{AT}^2$  is $\sim  |r|^2/w$; this  remains the
correct scaling form  for all $d>6$. Perturbation theory is in terms of the combination $x=w^2 |r|^{d/2-3}$.
 Hence the general form of the AT line, (at least for $6 <d <8 $), is 
\begin{equation}
h_{AT}^2 =\frac{|r|^2}{w} f(w^2 |r|^{d/2-3}),
\nonumber
\end{equation}
where $f(x)$ is a crossover function.  When $x$ is small, one is in the regime where perturbation theory is valid \cite{Green-Bray-Moore, Fisher-Sompolinsky}. In Ref. \cite{Moore-Bray} it was  shown that in this perturbative limit i.e. the limit where $T \rightarrow T_c$ at fixed $d$, $f(x) \sim (d-6)^4 x$ as $x \rightarrow 0$. The authors of Ref. \cite{Parisi-Temesvari} seem to have had a problem in understanding the existence of this  limit. However,  in order to investigate whether the AT line disappears as $d \rightarrow 6$ one needs the opposite limit, the limit when   $x$ is large or at least fixed. This is when $T $ is fixed (at a value less than $T_c$) or $|r|$ is set at some large value.  But this is the regime which is basically   non-perturbative and  no calculations valid  for it have been obtained: the only calculations which one can be done systematically are those which are perturbative.  It is this fact which makes determining the form of  the AT line in the limit  $d \rightarrow 6$  so  challenging.

Furthermore, the general problem of understanding how one can have a replica symmetric state for $d <6$ is also a non-perturbative task \cite{Moore}, which explains why analytical progress has been so slow.


\begin{references}
\bibitem{Parisi} G. Parisi, Phys.\ Rev.\ Lett. {\bf 43}, 1754 (1979); 
J. Phys.\ A {\bf 13}, 1101 (1980); {\em ibid.} {\bf 13}, 1887 (1980); Phys.\ 
Rev.\ Lett.\ {\bf 50}, 1946 (1983); M. M\'ezard, G. Parisi, N. Sourlas, 
G. Toulouse, and M. Virasoro, Phys.\ Rev.\ Lett.\ {\bf 52}, 1156 (1984).
\bibitem{Fisher-Huse} D. S. Fisher and D. A. Huse, Phys.\ Rev.\ Lett.\ 
{\bf 56}, 1601 (1986); Phys.\ Rev.\ B {\bf 38}, 386 (1988); {\em ibid.} 
{\bf 38}, 373 (1988).
\bibitem{Bray-Moore} A. J. Bray and M. A. Moore, Lecture Notes in Physics 
{\bf 275}, 121 (1986). 
\bibitem{McMillan} W. L. McMillan, Phys. Rev. B {\bf 29}, 4026 (1984).
\bibitem{AT} J. R. L. de Almeida and D. J. Thouless, J. Phys. A {\bf 11},
983 (1978).
\bibitem{Bray-Roberts} A. J. Bray and S. A Roberts, J. Phys.\ C {\bf 13}, 
5405 (1980).  
\bibitem{Moore-Bray} M. A. Moore and A. J. Bray, Phys. Rev. B {\bf 83}, 224408 (2011).
\bibitem{Moore} M. A. Moore, J. Phys. A {\bf 38}, L783 (2005).
\bibitem{Jorg} T. J\"{o}rg, H. G. Katzgraber, and F. Krzakala, Phys. Rev. Lett. {\bf 100}, 197202 (2008).
\bibitem{Janus} R. A. Ba\~{n}os et al., Proc. Natl. Acad. Sci. USA  {\bf 109}, 6452 (2012).
\bibitem{Lee-Young} L. W. Lee and A. P. Young,  Phys. Rev. E {\bf 72}, 036124 (2005).
\bibitem{Beyer-Weigel} F. Beyer and M. Weigel, Comp. Phys. Commun. {\bf 182}, 1883 (2011).
\bibitem{Aspelmeier-Moore} T. Aspelmeier and M. A. Moore, Phys.\ Rev.\ Lett. {\bf 92}, 077201 (2004).
\bibitem{Morris} B. W. Morris,  S. G. Colborne, M. A. Moore, A. J. Bray, and J. Canisius, J. Phys. C {\bf 19}, 1157 (1986).
\bibitem{Lee-Dhar-Young} L. W. Lee, A. Dhar and A. P. Young, Phys. Rev. E {\bf 71},  036146 (2005).
\bibitem{BWM} F. Beyer, M. Weigel, and M. A. Moore, cond-mat arXiv:1205.3975.
\bibitem{GBM} J. E. Green, A. J. Bray, and M. A. Moore, J. Phys. A {\bf 15}, 2307 (1982).
\bibitem{Harris-Lubensky-Chen} A. B. Harris, T. C. Lubensky, and J. H. Chen,
Phys.\ Rev.\ Lett.\ {\bf 36}, 415 (1976).
\bibitem{Viana} L. Viana, J. Phys. A {\bf 21}, 803 (1988).
\bibitem{Viana-Villarreal} L. Viana and C. Villarreal, J. Phys. A  {\bf 26}, 2873 (1993).
\bibitem{SY} A. Sharma and A. P. Young, Phys. Rev. E {\bf 81}, 061115 (2010).
\bibitem{Bray-Moore82} A. J. Bray and M. A. Moore, J. Phys. C {\bf 15}, L765 (1982).
\bibitem{AJKT} J. R. L. de Almeida, R. C. Jones, J. M. Kosterlitz, and D. J. Thouless, J. Phys. C {\bf 11}, L871 (1978).
\bibitem{KTJ} J. M. Kosterlitz, D. J. Thouless, and R. C. Jones, Phys. Rev. Lett. {\bf 36}, 1217 (1976).
\bibitem{1DAT} F. Beyer,  M. A. Moore, and M. Weigel, to be published.
\bibitem{Green-Bray-Moore} J. E. Green, A. J. Bray, and M. A. Moore,
J.  Phys. C {\bf 16}, L815 (1983). 
\bibitem{Fisher-Sompolinsky} D. S. Fisher and H. Sompolinsky, 
Phys.\ Rev.\ Lett.\ {\bf 54}, 1063 (1985). 
\bibitem{Parisi-Temesvari} G. Parisi and T. Temesvari,  Nucl. Phys.  B {\bf  858},  293 (2012).

\end{references}
\end{document}